\documentstyle[11pt,aaspp4,epsf]{article}

\def\etal   {{\rm ~et al.~}}
\def\kms    {\ifmmode{{\rm ~km~s}^{-1}}\else{~km~s$^{-1}$}\fi}
\def\lsun   {\ifmmode{{\rm ~L}_\odot}\else{~L$_\odot$}\fi}
\def\msun   {\ifmmode{{\rm ~M}_\odot}\else{~L$_\odot$}\fi}

\slugcomment{To appear in January 10, 1997 ApJ (Letters).}

\lefthead{Greenhill et al.}
\righthead{Water Maser Variability in Circinus}

\begin{document}

\title{Extremely Rapid Variations of \\ Water Maser Emission from 
the Circinus Galaxy}

\author{
L. J. Greenhill\altaffilmark{1}, 
S. P. Ellingsen\altaffilmark{2},
R. P. Norris\altaffilmark{3}, 
R. G. Gough\altaffilmark{3}, 
M. W. Sinclair\altaffilmark{3}, \\
J. M. Moran\altaffilmark{1},
R. Mushotzky\altaffilmark{4} }

\altaffiltext{1}{Harvard-Smithsonian Center for Astrophysics, 60 Garden St, 
Cambridge, MA 02138 USA,  greenhill@cfa.harvard.edu}

\altaffiltext{2}{University of Tasmania, G.P.O. 252C, Hobart, Tasmania, 
7001, Australia}

\altaffiltext{3}{Australia Telescope National Facility, CSIRO, P.O. Box 76,
Epping, NSW 2121, Australia}

\altaffiltext{4}{NASA/Goddard Space Flight Center, Laboratory 
for High Energy Astrophysics, Code 666, Greenbelt, MD 20771 USA}

\begin{abstract}

The water maser lines in the Seyfert nucleus of the Circinus galaxy 
vary on
time scales as short as a few minutes. The amplitude of one line 
more than doubled in $\approx 10$ minutes, reaching $\approx 37$ Jy, which 
corresponded to an increase of $\approx 6$ L$_\odot$, assuming isotropic 
emission, in a single maser feature on a {\sl size} scale of about 1 AU, 
based on
light-travel time. Other lines vary by 
up to about 30\% on similar time scales.
The variability is at least two orders of magnitude more rapid than any
observed for other Galactic or extragalactic water masers. The intensity
changes cannot be attributed easily to a mechanism of intrinsic fluctuations.
The variability may be the result of strong interstellar diffractive
scintillation along the line of sight within our Galaxy. This would be the
first example of diffractive scintillation for any source at 22 GHz and for
any source other than a pulsar. However, only the very shortest timescales for
interstellar scintillation, obtained from pulsar observations and scaled to 22
GHz, correspond to the observed maser variability.
Alternatively, the intensity changes may
be a reaction to fluctuations in compact background or radiative pump sources 
and thereby may be related to variability of the central engine.

The maser spectral features symmetrically bracket the systemic velocity of the
galaxy, with components red- and blue-shifted by about $\pm(100-200)\kms$. The
spectrum of the Circinus maser is similar in some respects to that of the
maser in NGC~4258, which probably traces a molecular disk rotating around a
supermassive object. VLBI observations could reveal whether the maser
source in the heart of the Circinus galaxy is part of a similar dynamical
system.

\end{abstract}


\section{Introduction} 

Renewed interest in extragalactic water masers has been stimulated by
observations of the maser in the nucleus of NGC 4258 (\cite{Miy95}). Maser
emission is present at the systemic velocity of the galaxy and in satellite
lines offset by about $\pm 900$\kms. The emission arises from a thin, nearly
edge-on molecular disk that is in Keplerian rotation around a massive central
object (see also \cite{M95}).  These studies
have established, to high accuracy, the binding mass and the geometry of this
system on sub-parsec scales.

The Circinus galaxy (CG) is a Seyfert 2 galaxy at a distance of about 4 Mpc
that has long been known to harbor a water maser in its nucleus (\cite{GW82}).
The nucleus shows a one-sided ionization cone (\cite{Mar94}) and bipolar radio
lobes that extend about 200 pc along the minor axis of the galaxy
(\cite{Elm95}). Excitation of coronal lines within the inner 10 pc is
consistent with a photoionizing central continuum source, rather than with a
starburst, though there are signs of an old starburst within the innermost 4
pc (\cite{Oliva94}). A heavily obscured hard (2--10 keV) X-ray source
(\cite{Matt96}) and a compact radio synchrotron source also lie in the nucleus
(\cite{Nor96}).

The water maser spectrum of the CG is similar to that of NGC~4258 in two ways.
First, there are two complexes of narrow lines that bracket the systemic
velocity of the galaxy. Second, there is a distinct complex of emission lines
close to the systemic heliocentric velocity of 416\kms~(\cite{LH95}). (We use
the radio definition of radial velocity throughout.) The red-shifted lines in the
CG were discovered by Gardner \& Whiteoak (1982), but the blue-shifted lines
have only recently been detected by us (reported in this paper),
and by Nakai\etal (1995) and Braatz, Wilson, \& Henkel (1996), independently.

Whiteoak \& Gardner (1986) observed that the shape of the CG maser spectrum
varies substantially in about one month. The relative intensities of spectral
components in the NGC~4258 maser can also vary by tens of percent in a few
weeks (\cite{Gre95drift}). The maser in IC~10 can vary by 60\% in less than
one day (\cite{argon94}), which may be a consequence of interstellar
scattering. We note that the most rapid known flares in Galactic interstellar
H$_2$O masers have time scales of one to ten days (e.g., \cite{Lil93};
\cite{PS93}; \cite{Has77}). In this {\sl Letter}, we report the measurement of
a finely sampled time-series of spectra, which has revealed variability on
time scales at least {\sl two} orders of magnitude shorter than those known
for any other water maser. We also analyze the morphology of these spectra in
the context of an edge-on disk geometry.

\section{Observations} 

The observations were carried out at the Australia Telescope National Facility's
Parkes radio telescope in 1995 October, as part of a search for new water maser
sources in 28 galaxies ($z<0.1$) with high X-ray absorption columns or excess
infrared emission accompanied by compact radio continuum emission. We used a
newly commissioned 22 GHz receiver, with two orthogonal linear polarization
channels. From a tipping scan, we estimate that the receiver temperature was 95
K, and that the atmospheric opacity at zenith was 0.06. The telescope
sensitivity was about 5.7 Jy K$^{-1}$, based on a flux density of 21.2 Jy for
Virgo A. Uncertainties in pointing are responsible for up to 10\% loss in gain.
We adopted the gain curve of Bourke (1994), increased by about 10\% at low
elevation angles as indicated by our observations of Virgo A. We observed in
position-switching mode, with four 16 MHz bandpasses tuned to cover the range of
emission from 195--735\kms. The 4096-channel digital autocorrelator provided a
channel spacing of 15.625 kHz, or 0.21\kms.

\section{The observed spectrum and variability} 

The spectrum of the CG water maser shows two complexes of strong lines between
510 and 610\kms, and between 240 and 310\kms~(see Fig.~1). We refer to these as
the red and the blue satellite lines. Spectra taken in previous years show
similar velocity ranges for the complexes of satellite lines (\cite{WG86};
\cite{Nak95}). Figure 2 shows the relative flux density variations for the
three strongest maser lines on time scales as short as a few minutes. The
567.25\kms~feature more than doubled in strength to 
about 37 Jy in only 10 minutes between 15 and 16
hours LMST on day 289. On the assumption of isotropic emission
this represents an increase from 4.8 to 11 L$_\odot$. At other times
the strength of this feature varies by about 50\%, while the other two features 
vary upto about 30\%, on similar time scales.
We conservatively adopt a characteristic time scale for variability,
$t_{obs}$, of $\sim 300$ s.

\begin{figure}[ht]
\plotfiddle{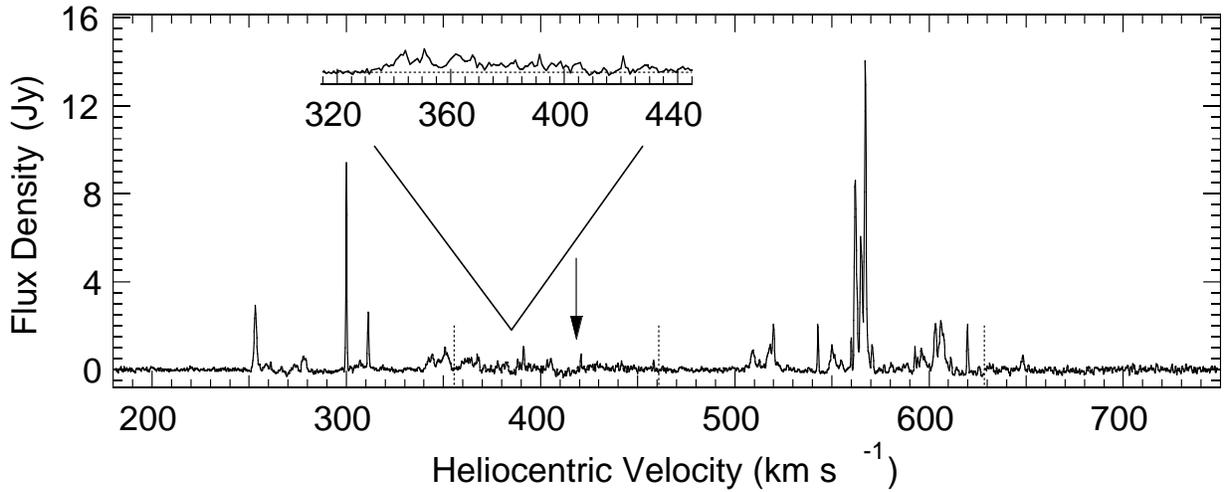}{2.0truein}{0}{90}{90}{-265}{-230}
\caption{
Time-averaged spectrum of the water maser emission from the nucleus of
the Circinus galaxy. The heliocentric systemic velocity of the galaxy, 416\kms,
is indicated by the arrow. The dashed lines indicate the divisions between the
four receiver bandpasses. The on-source integration time for the two bandpasses
that contain the two strong complexes of lines was 350 minutes ($1\sigma$ noise =
0.05 Jy). The integration time for the other bands was 100 minutes ($1\sigma$
noise = 0.1 Jy). The spectrum was Hanning smoothed and has a channel spacing of
$0.21\kms$. The three spectral features examined in Figure 2 are indicated by
the bars along the top axis. ({\it insert}) Spectrum obtained separately that
shows more clearly the emission near the systemic velocity (as a consequence of
better baseline removal). The on-source integration time was 60 minutes and the
channel spacing was 0.8\kms. The gain calibration is accurate to $\approx 10$ \%
and includes corrections for atmospheric attenuation, antenna gain variations,
and average pointing error.
}
\end{figure}

We feel certain that instrumental effects are not responsible for the apparent
maser fluctuations. The variations of different spectral features are not
correlated. Moreover, observations of the CG and NGC~4945 masers were made
within hours of each other, and no variability was detected in NGC~4945.
Apparent variability might be a consequence of telescope pointing errors if the
maser features were distributed over an area comparable to the $1\rlap{.}'3$
primary beam. However, spectra taken at offset positions show that the region of
maser emission is relatively compact, $< 0\rlap{.}'3$, as expected for an
extragalactic nuclear maser. False variability might also arise from
under-sampling of the spectral lines but in our case the half-power widths
(FWHM) of the maser features are at least four spectral channels.

\begin{figure}[ht]
\plotfiddle{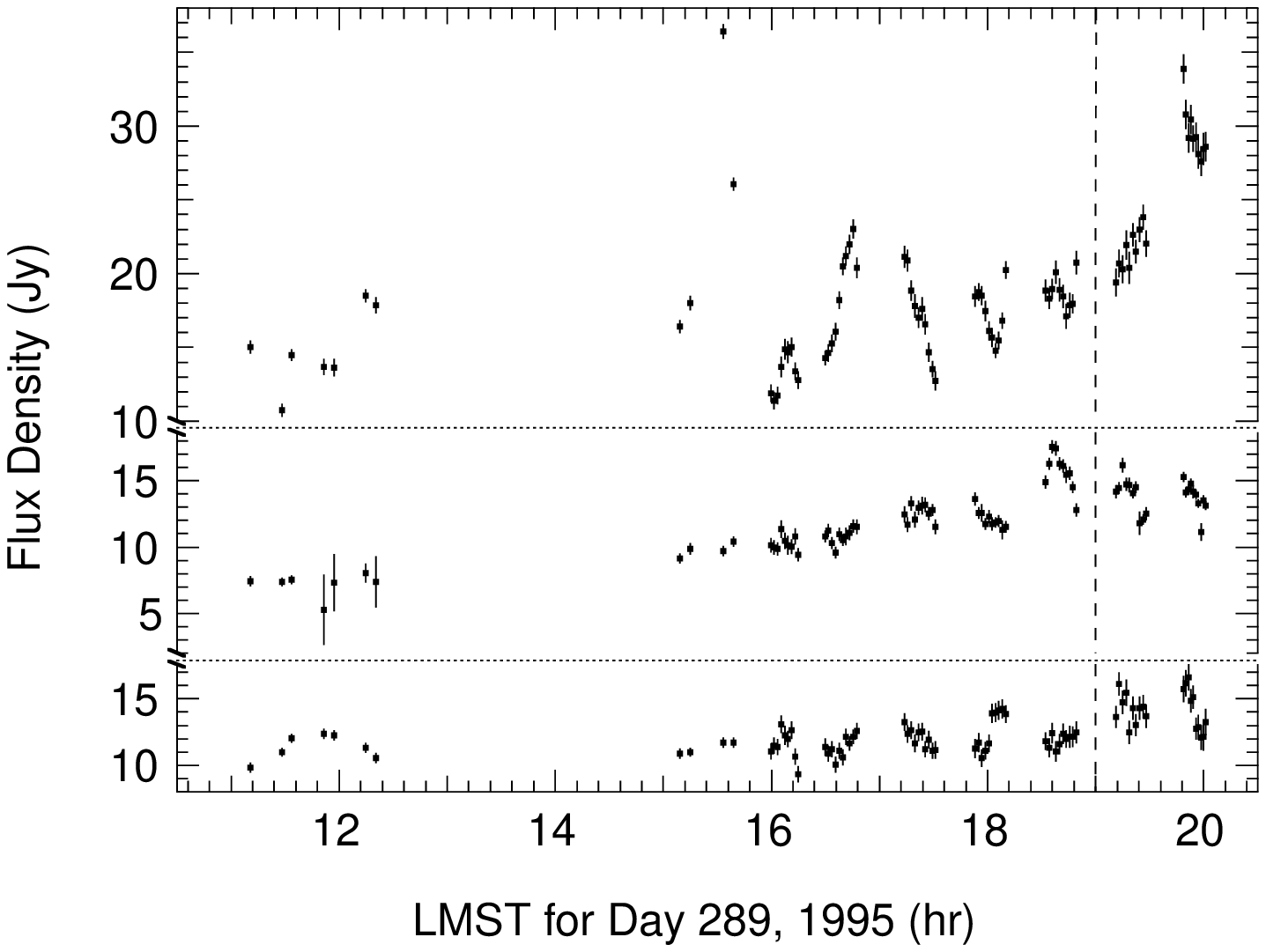}{4.0truein}{0}{100}{100}{-290}{-180}
\caption{
Radio light curves for three of the strongest spectral components, ({\it
top}) 567.25, ({\it middle}) 562.00, and ({\it bottom}) 299.91\kms. Instrumental
effects cannot be responsible for the time variability. Before 16 LMST data were
taken in 5 minute integrations, and after 16 LMST in 1 minute integrations. Data
gaps occurred when the telescope was off source.  The error bars reflect the
formal uncertainties from fitting Gaussian line profiles to the spectra. It is
possible that an error in the calibration at the lowest elevations causes a 
slow upward trend in flux densities after 19 LMST, on the order of 10\%.
}
\end{figure}

The FWHM of the narrowest satellite spectral line is about 0.9\kms. The
satellite lines seen in NGC~4258 are about 1.0\kms~wide (\cite{Miy95}), and the
narrowest water maser lines ever observed are about 0.3\kms~in IC~10
(\cite{argon94}). Emission from masers in the presence of a magnetic field may
be linearly polarized (\cite{WWD94a}). We used our dual polarization
observations of the satellite lines to construct Stokes Q spectra of the
source. There is no statistically significant linear polarization at a level of 
0.5 Jy  (4\% for a 12 Jy line).

\section{Origins of Variability}

The maser variability is extraordinary because a light travel time of a few
minutes implies a characteristic scale of about 1 AU. Masers may be visualized
typically as filamentary structures, viewed end-on, which mark local
enhancements in density, velocity coherence, or excitation conditions
(\cite{Elit92}). Large fluctuations in emission rate are caused by changes 
in the maser medium over lengths greater than the maser gain length, $\ell$,
which is the characteristic length over which the line-center opacity changes
by $\sim 1$. For typical conditions, $\ell \gg 1$ AU (e.g., a maser pump
efficiency of 1\%, a fractional water abundance of $10^{-4}$, and a temperature of
400 K as in \cite{RM88}). To be responsible for the observed maser
variability, changes in the maser medium must propagate at about the speed of
light, and we know of no suitable driving mechanism.

There are three mechanisms by which the maser may vary in response to processes
external to the maser medium: interstellar scintillation, amplification of a
time-variable background source, and excitation by a time-variable radiative
pump source. The last two mechanisms assume the presence of a compact (1 AU)
source in the vicinity of the maser. Large amplitude fluctuations on time scales
close to $10^3$ s have been observed towards nuclei at X-ray energies
(\cite{MDP93}), as in NGC~4051 (\cite{Guain96}) and Mrk~766 (\cite{Lei96}).

Interstellar scintillation (ISS) is caused by scattering in an ionized
component of the Galactic interstellar medium, which may be important because
the CG lies at a Galactic latitude of only $-3.8^\circ$ (longitude
$311^\circ$). We consider scintillation caused by a thin, turbulent scattering
screen at distance $D$ that is moving with respect to the observer at velocity
$v$. The turbulence is usually described by a Kolmogorov power law
(\cite{Arm81}) and the scintillation is divided into two regimes, weak and
strong. For a point source, the time scale of weak ISS is approximately the
crossing time of the Fresnel scale, $\ell_F\approx\sqrt{D\lambda/2\pi}$, where
$\lambda$ is wavelength (\cite{Nar93}). For $D=1$ kpc, $\lambda=1.35$ cm, and
$v=100\kms$, $\ell_F\approx 2.5\times10^{10}$ cm and $t_F=\ell_F/v\approx 4000$ s
(\cite{Rick90}), about an order of magnitude greater than $t_{obs}$.  For
strong ISS the contributing turbulent length scales obey $\ell_{scin} < \ell_F$.
(Strong scintillation may be diffractive or refractive. We do not consider the
second because the relevant length scales are longer.) Diffractive
scintillation at 1 GHz with time scales of $t_{scin}>20$ s is observed toward
pulsars for which $D> 1$ kpc and $2^\circ<b<5^\circ$ (\cite{Cord86}). For
Kolmogorov turbulence, $t_{scin} \propto\nu^{1.2}$, where $\nu$ is frequency.
Goodman \& Narayan (1985) show that for 
non-Kolmogorov scattering screens $t_{scin}\propto\nu^{1.0}$. In either case,
only the fastest
observed scintillation at 1 GHz, $t_{scin}\sim 10$ s, corresponds to
sufficiently rapid variations at 22 GHz. However, the true angular size of a
scintillating maser feature, $\theta_m$, must also be less than
$\theta_{scin}=\ell_{scin}/D$ because otherwise the scintillation is slowed by a factor
of $(\theta_m/\theta_{scin})$ and the amplitude is similarly decreased
(\cite{Nar93}). Hence, for strong scintillation, $\theta_m < \theta_{scin} <
\theta_F$, which corresponds to $<7$ AU at a distance of 4 Mpc, where
$\theta_F=\ell_F/D$.  The scintillation time scale for a scattering screen in
the AGN, within $< 1$ kpc of the maser, may be $< t_{obs}$, but the amplitude
of the variations will be substantially reduced because the angular size of
the maser source, viewed from the screen, will be large.

The time scale of variability for a maser that amplifies a background source
is not limited by the extent of the maser. The maser medium forms an image of
the background source, and the angular sizes of the two are the same (see
\cite{Has90}). For the CG, the background source must be less than a few
light-minutes in size, and the apparent maser cross-section would be
about 0.25 $\mu$as or 1 AU. We note that substantial variability may be
possible only if a fraction of the maser is unsaturated since the emission
rate of a completely saturated maser is fixed by the pump rate and beam angle.
In this case, the beam angle is set by the solid angle of the background
source as seen from the maser (see \cite{Has90}).

The variability time scale of a radiatively pumped maser is also independent
of its physical extent, apart from optical-depth effects. The pump luminosity,
which may be supplied directly by the central engine, must be large enough
that the pump rate exceeds the maser emission rate. For Galactic interstellar
water masers, radiative pumping by stellar radiation generally cannot support
observed emission rates however, substantially higher luminosities are present
in AGN. For a single pump transition (at wavelength $\lambda_p$), $F_{\nu
p}>F_{\nu m}(\Omega_b/\Omega_m)(x_m/x_p)$, where $F_{\nu p}$ and $F_{\nu m}$
are the pump and maser flux densities, respectively, $\Omega_b$ is the maser
beam solid angle, $\Omega_m$ is the solid angle of the maser viewed from the
pump, and $x_p$ and $x_m$ are the fractional bandwidths of the pump and maser,
which are similar (\cite{RM88}). For $N$ simultaneous pump transitions the requisite
$F_{\nu p}$ is reduced proportionately. We assume the central engine has a
flat spectral energy distribution ($\nu F_\nu$) between ultraviolet and X-ray
energies and estimate $F_{\nu p}$ at $\lambda_p$ from the inferred 
intrinsic X-ray luminosity of the central engine. Matt\etal (1996) obtain a
luminosity of $4\times 10^{41}f^{-1}$ erg s$^{-1}$ from the observed X-ray
spectrum, assuming that a fraction, $f=(\Omega_R / 2\pi)$, 
is reflected by a portion of the obscuring circumnuclear
torus into the line of sight, where $\Omega_R$ is the solid angle of the
reflecting material seen from the central engine. Since the infrared
luminosity of the nucleus is $\approx 5\times10^{43}$ erg s$^{-1}$
(\cite{MG84}), $f$ is on the order of a few percent. Based on order of 
magnitude estimates for the parameters involved, radiative pumping requires
that $\lambda_p \sim 0.6 N^{-1} ({\Omega_b\over\Omega_m}) ({F_{\nu m}\over 10 {\rm Jy}}) ({f\over 0.01}) 
({D\over 4~{\rm Mpc}})^2 $, where $\lambda_p$ is
expressed in units of microns, and $({\Omega_b\over\Omega_m})\sim 1$
(\cite{RM88}). Hence, a radiative maser pump may be feasible if it consists
of on the order of 10 transitions at ultraviolet wavelengths.

Several other specific mechanisms for maser variability have been proposed in
the literature, though none of them explains the observed time scale. First,
in partially saturated masers, a radiative instability  can cause variations
with a period of $L/c$, where $L$ is the maser filament length (\cite{SW92}).
However, strong maser emission requires many gain lengths or $L/c\gg t_{obs}$.
 Second, masers may amplify each other, and an outburst can occur when one
maser drifts in front of another with the same line-of-sight velocity
(\cite{DW89}). Third, rotation of a maser beam across the line of sight can
mimic rapid intensity variations (\cite{NB81}). However, for the latter two
mechanisms, maser lengths of $\gg1$ AU and transverse velocities of $<100\kms$
require maser cross-sections of $<10^{10}$ cm, which are improbably small.

\section{Structure of the maser source}

Based on the morphology of the H$_2$O maser spectrum, we propose that the maser
emission arises from a nearly edge-on Keplerian disk, as in NGC~4258. The
satellite lines arise on a disk diameter perpendicular to the line of sight, and
the systemic emission arises on the near side of the disk, close to the line of
sight to the dynamical center. These locations naturally possess the greatest
line-of-sight velocity coherence. For Keplerian rotation, the outer radius of
the disk is about four times the inner radius because the velocities of the
satellite lines at these radii, relative to the systemic velocity, differ by a
factor of two (i.e., 100 and 200\kms, respectively). The binding 
mass for the disk in solar units is $10^6r_{0.1}\sin^2i$,
where $r_{0.1}$ is the disk inner radius in units of 0.1 pc, and $i$ is the
inclination of the disk rotation axis. A  central engine of $<10^7$ M$_\odot$,
with a gravitational radius of $<3\times10^{12}$ cm, can vary on time scales of
100 s (cf., Mushotzky\etal (1993)), as may be required by the observed maser variability.  We note
that for rotation speeds of 200\kms~and $r_{0.1}>0.04$,  the line-of-sight centripetal
acceleration of systemic maser features in the spectrum will be $<1$\kms yr$^{-1}$ 
(cf., \cite{Gre95drift}). 

\section{Conclusions}

We have observed extremely rapid variability in the spectrum of the nuclear
water maser in the Circinus galaxy. In principle, fluctuations on time scales
of a few minutes may be explained by ISS in our Galaxy,
variability of background emission that is being amplified, or variability of
a radiative pump. While only the shortest measured scintillation time scales
for pulsars, extrapolated to 22 GHz, are acceptable, the sensitivity limits of 
the pulsar observations may hide 
still more rapid diffractive ISS. In addition, the scattering medium at low latitude is 
clumpy, which may also allow more rapid scintillation along the 
particular line of sight to the CG. If the maser fluctuations are caused by diffractive
ISS, then the maser is the first example for any source at
22 GHz and for any source other than a pulsar.  We note that the only other
extragalactic water maser source reported to display significant variations
over $<1$ day (IC~10) also lies at low latitude, $3^\circ$ (\cite{argon94}).

Since the maser spectrum consists of independently varying components, the
alternative background amplification and radiative pump schemes require either
several compact fluctuating sources or variable pump illumination that depends
on location or Doppler velocity. Nonetheless, if the CG maser directly
reflects the intrinsic variability of the central engine on time scales of a
few minutes, then the maser will be a powerful probe of this heavily
obscured AGN. Rapid variability has been observed exclusively in X-ray 
wavebands for other AGN but this does not necessarily preclude the possibility
that emission from central engines also varies about as rapidly 
at other wavelengths.

Scintillation and background amplification models place stringent upper limits
of 1 -- 10 AU on the apparent sizes of the maser components, which implies
very large brightness temperatures, $>10^{16}$ K. The general symmetry of the
maser spectrum with respect to the systemic velocity is similar to that of
NGC~4258, in which the masers lie in a Keplerian molecular disk bound by a
massive central engine. In the absence of VLBI data, we propose that the same
model may be applicable to the CG maser. The highly ordered velocity field of
a thin Keplerian disk naturally supports highly elongated masers, which Watson
(1994b) argues is a necessary condition to permit the inferred high brightness
temperatures

\acknowledgments

We thank D. Backer, C. Gwinn, R. Manchester, R. Narayan, and W. Watson, for 
helpful discussions. We are also grateful to Euan Troup, Harry Fagg, and Warwick
Wilson for their special assistance at Parkes.

\end{document}